\documentclass{osa-article}
\journal{oe}

\usepackage{lineno}
\usepackage{subfigure}
\usepackage{multirow}
\usepackage{indentfirst}
\usepackage{color, colortbl}

\begin{document}

\title{High Density Vertical Optical Interconnects for Passive Assembly}

\author{Drew Weninger,\authormark{1, 4}, Samuel Serna, \authormark{3} Achint Jain, \authormark{1,4}, Lionel Kimerling\authormark{1,4}, and Anuradha Agarwal,\authormark{2,4}}

\address{\authormark{1} Materials Science and Engineering Department\\
\authormark{2} Materials Research Laboratory\\
\authormark{3}Bridgewater State University, Physics Department, 131 Summer St, Bridgewater, MA 02324, USA\\
\authormark{4}Massachusetts Institute of Technology, 77 Massachusetts Avenue, Cambridge, MA 02139, USA}

\email{\authormark{*}drewski@mit.edu}


\begin{abstract*}
The co-packaging of optics and electronics provides a potential path forward to achieving beyond 50 Tbps top of rack switch packages. In a co-packaged design, the scaling of bandwidth, cost, and energy is governed by the number of optical transceivers (TxRx) per package as opposed to transistor shrink. Due to the large footprint of optical components relative to their electronic counterparts, the vertical stacking of optical TxRx chips in a co-packaged optics design will become a necessity. As a result, development of efficient, dense, and wide alignment tolerance chip-to-chip optical couplers will be an enabling technology for continued TxRx scaling. In this paper, we propose a novel scheme to vertically couple into standard 220 nm silicon on insulator waveguides from 220 nm silicon nitride on glass waveguides using overlapping, inverse double tapers. Simulation results using Lumerical's 3D Finite Difference Time Domain solver solver are presented, demonstrating insertion losses below -0.13 dB for an inter-chip spacing of 1 $\mu$m, 1 dB vertical and lateral alignment tolerances of approximately $\pm$ 2.7 $\mu$m, a greater than 300 nm 1 dB bandwidth, and 1 dB twist and tilt tolerance of approximately 2.3 degrees and 0.4 degrees, respectively. These results demonstrate the potential of our coupler for use in co-packaged designs requiring high performance, high density, CMOS compatible out of plane optical connections.

\end{abstract*}

\section{Introduction}
In 2021, 20.6 ZB of the global datasphere - or near 30\% - traveled through data centers, where over 85\% of that data remained locally in "East-West" traffic \cite{2018cisco}. In current data center architectures, information is routed to servers within a rack by using an Ethernet switch located at the top of the rack (ToR). 
In the past, ToR switch packages have met bandwidth challenges like this by doubling total bandwidth capacity every two years. This was accomplished either by doubling the number of copper (Cu) input/output (I/O) channels in parallel to the package (which occurred every two years) or by doubling the bandwidth capacity per Cu channel (which occurred every four years) \cite{2021minkenburg}. High performance ToR switch packages have been demonstrated using all electrical I/O with 512 channels, each operating at 106.25 Gbps per channel using the IEEE 802.3-ck standard, yielding a total of 51.2 Tbps per package \cite{2021fathololoumi}. 
\par Scaling beyond this point using Cu I/O has many challenges, an important one being that the crosstalk between Cu lines increases with increasing frequency and increases as more laminates are squeezed into the same package substrate thickness, which is already at or above 12 layers \cite{2016lau}. 

On the other hand, all-optical I/O, which does not suffer from the skin effect, becomes energetically favorable over Cu I/O for distances longer than 1 mm when using data rates above 100 Gbps \cite{2018thraskias}. On top of that, all-optical I/O can scale in bandwidth capacity through increasing the number of wavelengths per channel in wavelength division multiplexing schemes since in a linear material different wavelengths do not interact with one another. Thus, a sustainable solution is to implement "co-packaged optics" by placing the transceiver (TxRx), the device used to convert optical signals to electrical signals and vice versa, within the switch package to minimize Cu length, and to use all-optical I/O at the package interface. 

\begin{figure}[htbp]
\centering
\includegraphics[scale=0.4]{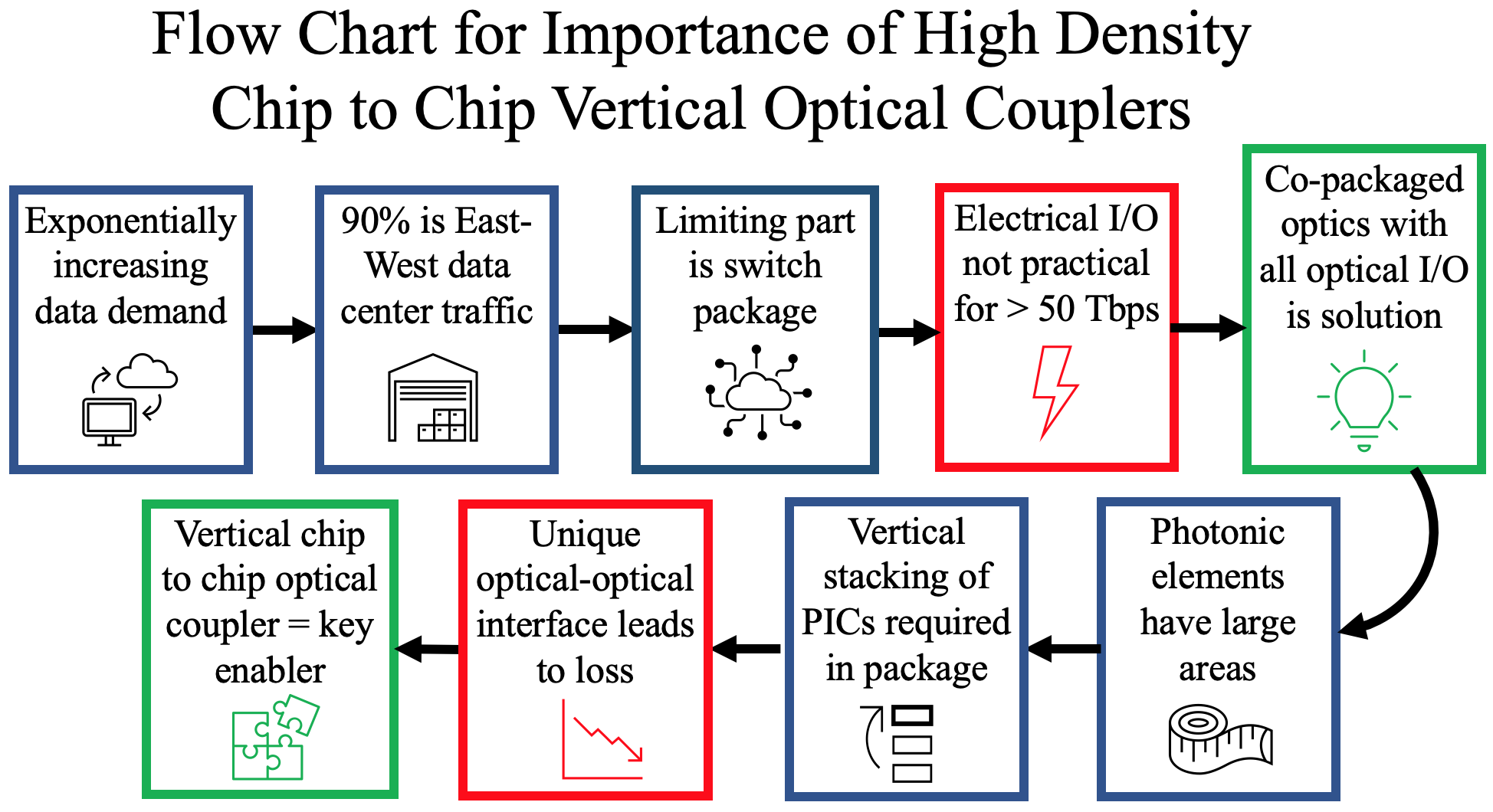}
\caption{A flow chart for understanding how high density chip to chip vertical optical couplers will be a key enabling technology for meeting global data demands. The red boxes highlight problems which arise in the flow and the green boxes help highlight key solutions for industry to converge towards. In other words, the flow chart is meant to emphasize the train of thought which shows how all-electrical I/O is not scalable in terms of bandwidth capacity, energy efficiency, or cost per package for greater than 50 Tbps ToR switch packages, and how co-packaged optics utilizing vertical optical couplers is a feasible, sustainable solution.}
\label{fig:logic_flow_chart}
\end{figure}

\par A first step towards achieving greater than 50 Tbps ToR switch packages utilizing all-optical I/O is the development of TxRx modules which are able to be co-packaged with electrical devices. Several different co-packaged TxRx designs have experimentally shown bandwidths of 1.6 to 2.0 Tbps operating with TxRx energy efficiencies between 20 pJ/bit and 4.9 pJ/bit, respectively \cite{2020sun, 2021fathololoumi}. Other similar co-packaged TxRx have been simulated and show a bandwidth capacity of 3.58 Tbps with an energy consumption of less than 2 pJ/bit per TxRx \cite{2021maniotis}. An important aspect of these co-packaged designs is that scaling of bandwidth capacity, energy efficiency, and cost per package is no longer associated with transistor shrink, but instead with the number of TxRx per package. As more TxRx are located inside the package to improve performance, the available package area will become limited and the stacking of photonic integrated circuits (PICs) in the vertical direction on optical interposers or directly on other PICs will become a necessity. This is especially true because optical components such as micro-ring resonators (MRRs), Mach-Zehnder interferometers (MZIs), and photo-detectors (PDs) take up a significantly larger area than their electronic counterparts (and thus, PICs take up larger areas than electronic ICs). However, by vertically stacking optical chips, another optical-optical interface is created which will contribute to the loss budget of an optical network. Therefore, an innovation in out of plane couplers that can vertically route light from one waveguide to another waveguide on a separate chip with high efficiency and with a dense pitch will be an enabling technology for co-packaging of optics and electronics. A flow chart summarizing the line of thought which arrived at this conclusion is outlined in Figure \ref{fig:logic_flow_chart}, with important problems highlighted in red and important solutions highlighted in green.

\par
\begin{figure}[htbp]
\centering
\includegraphics[scale=0.55]{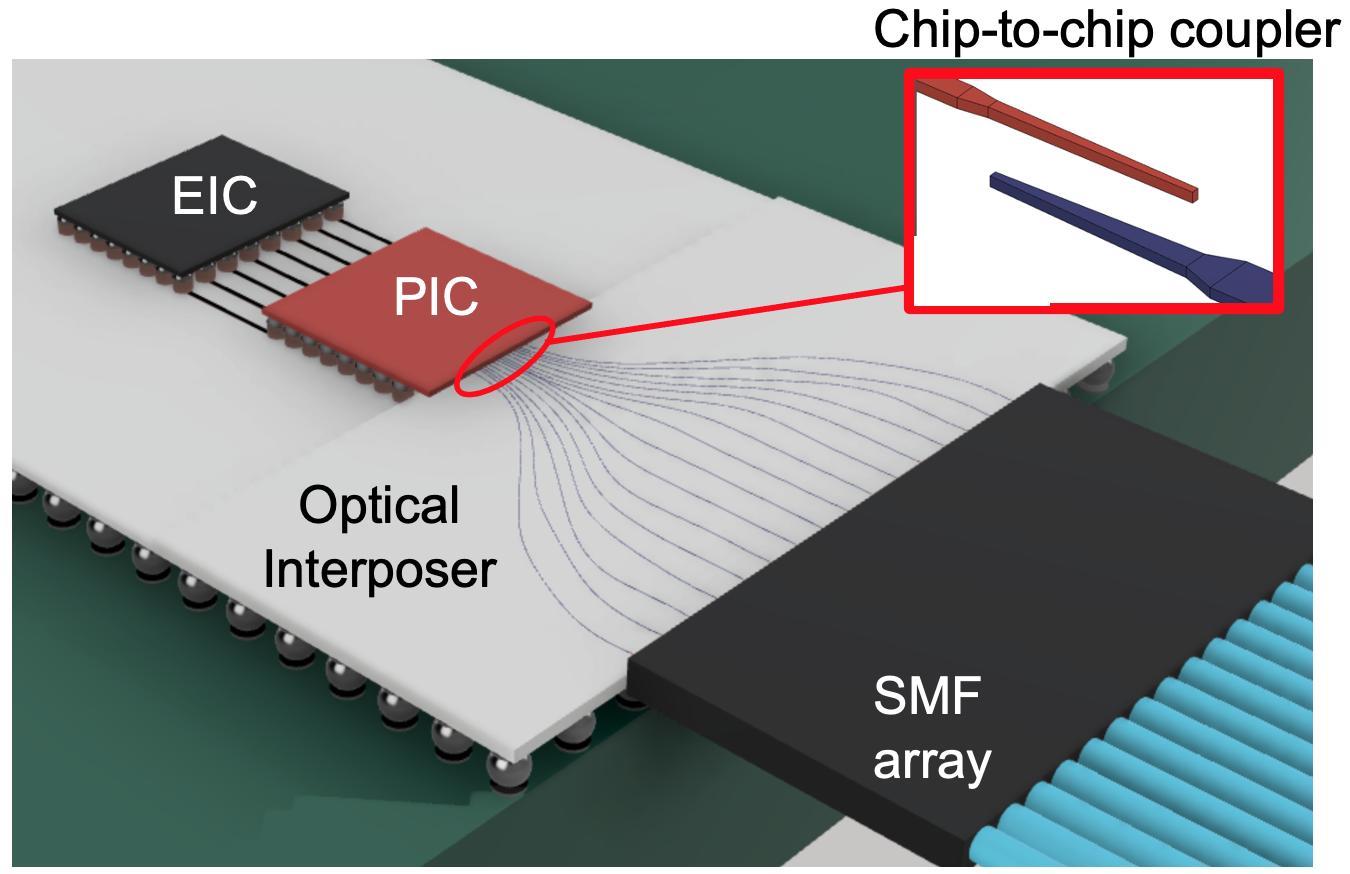}
\caption{A schematic of the proposed co-packaged design which uses high density vertical optical interconnects to enable optical fanout to an array of SMFs is shown. The design is composed of a SiO$_2$ optical interposer shown in white with an EIC shown in black and a silicon PIC shown in red, both of which sit on top of the interposer and are assembled using pick and place technology. Note that while only one PIC was shown in this diagram, it is intended to hold a large number of chips, and that the diagram is not to scale in terms of electronic or photonic chip sizes. 
}
\label{fig:optical_fanout}
\end{figure}

 The vertical optical interconnect design [\textbf{US Patent No. 11067754}] presented in this paper provides a solution with low insertion loss, high translational and rotational alignment tolerance, CMOS process flow compatibility, and fine lateral pitch. In fact, it is only one aspect of a broader 2.5D co-packaged optics design, which is shown in Figure \ref{fig:optical_fanout}. In this context, 2.5D refers to the placing of PICs and EICs side by side as opposed to stacking EICs on top of PICs in a 3D configuration. The diagram shows an SiO$_2$ optical interposer providing optical fanout from the TxRx to an array of SMFs. By implementing optical fanout, this design allows for a significantly higher density of optical I/O along the available shoreline of the PIC (the TxRx) because the SMF pitch is limited to 125 $\mu$m by the fiber cladding, but the waveguides on the optical interposer are limited only by the mode size of a single mode waveguide. Additionally, the use of an optical interposer allows for high speed pick and place tools to be used for assembly of PICs instead of using flyover fiber arrays attaching to each TxRx individually \cite{samtec}.  Development of three enabling technologies make this design possible: a novel chip to chip evanescent coupler, a novel SMF to interposer expanded beam coupler, and Cu micro-pillar arrays which provide dense electrical connections to the PIC from the EIC (e.g. 60 $\mu$m pitch size or less) and passive self-alignment of the vertical coupler elements after pick and place of the PIC is completed. The passive self-alignment occurs due to the surface tension of the molten Sn-Ag-Cu (SAC) cap which sits atop the Cu $\mu$-pillar joint. The experimental self-alignment of Cu $\mu$-pillars (also called C2 bumps) and standard ball grid array joints (called C4 bumps) has been shown on numerous occasions to exhibit a final misalignment of less than 1 $\mu$m following bonding \cite{1990wale, 1992hayashi, 2005bernabe, 2006hutter, 2012kong, 2012bernabe, 2015nah, 2016lee, 2017zonou, 2018park}. Through this assembly mechanism, the need for active alignment of optical components can be significantly reduced or eliminated. This is crucial because active alignment increases fabrication time and the overall package cost, which already accounts for over 80\% of the total cost for photonic packages \cite{2020IPSRpackaging}. The following sections focus on describing the design and performance of our vertical coupler because it is the most challenging and important aspect of the proposed package design.

\subsection{High Density Vertical Optical Interconnects}
\begin{figure}[htbp]
\centering
\subfigure[\label{subfig:perspectiveview}]{\includegraphics[scale=0.6]{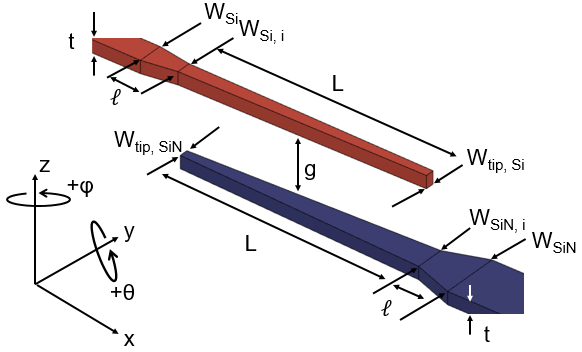}}
\subfigure[\label{subfig:topview}]{\includegraphics[scale=0.4]{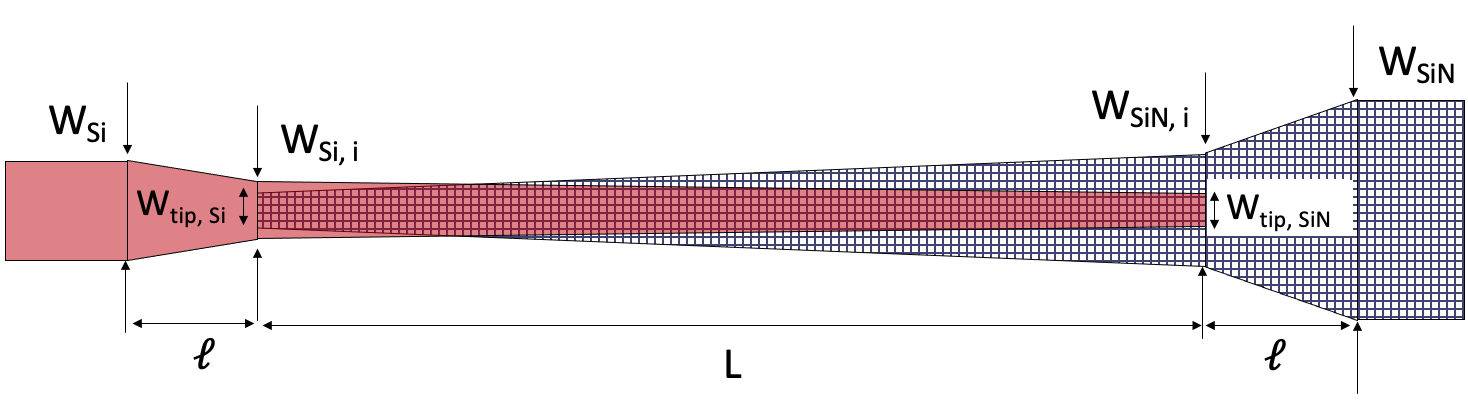}}
\subfigure[\label{subfig:sideview}]{\includegraphics[scale=0.35]{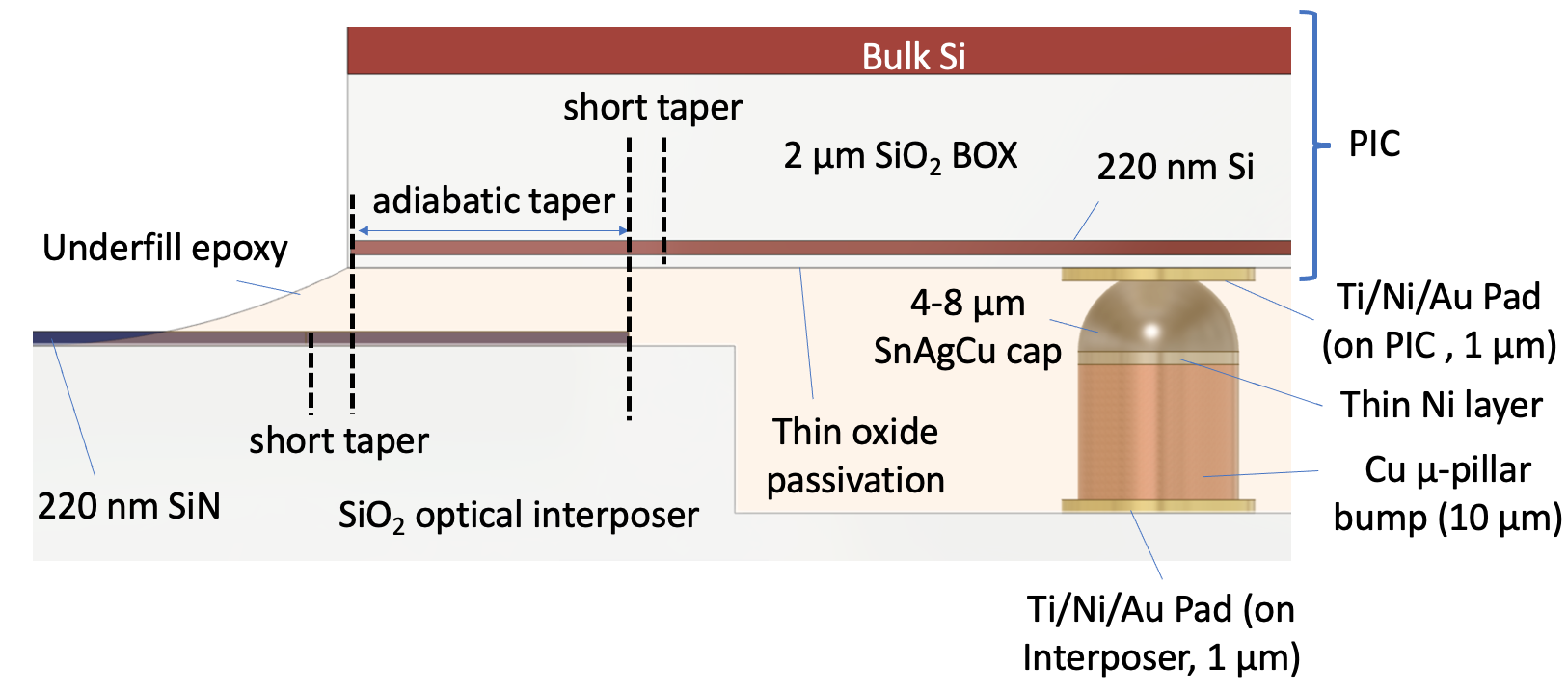}}
\caption{The basic design of our novel vertical optical interconnect. In \subref{subfig:perspectiveview}, a perspective view highlights the fundamental design parameters of our structure. The lower taper is shown to be a different color because it is fabricated out of SiN as opposed to Si. Also, the coordinate system used for the remainder of the study is shown, with $z$ pointing vertically, $y$ pointing laterally, $x$ longitudinally, $+\phi$ showing counterclockwise rotation around the $z$ axis (representing twist rotation), and $+\theta$ showing rotation around the $y$ axis (representing tilt rotation). In \subref{subfig:topview}, a top view of the coupler is shown. Notice that the adiabatic taper lengths (L) of the upper and lower tapers are equal and the tapers overlap such that the tip of the Si adiabatic taper ends at the $x$ point where the SiN adiabatic taper begins (when in perfect alignment). In \subref{subfig:sideview}, a side view of the structure (not to scale) after the PIC has been bonded to the interposer is shown with different colors to help show the material stack. Here, the C2 bumps are deposited into a trench etched using deep RIE. 
}
\label{fig:crosstapers}
\end{figure}

The basis for our vertical evanescent coupler design is two crossed (overlapping), inverse, double tapers as shown in Figure \ref{fig:crosstapers}. Different variations of this design were developed in \cite{2008rong, 2016itoh, 2017itoh, 2019macfarlane, 2015gao, 2013jones, 2017sacher, 2020bai}. Our design is distinct from these in several ways. First, it is intended for coupling between two separate chips (inter-chip coupling) as opposed to coupling within two layers of the same chip (intra-chip coupling). Second, our design uses standard 220 nm silicon nitride (SiN) for the material of the waveguide and double taper on the lower chip, which has multiple advantages. For one, it leads to a reduction in propagation loss on the interposer due to the lower refractive index of the SiN compared to a design where the waveguides on the interposer are fabricated from 220 nm Si. In addition, using standard 220 nm SiN for the lower taper and 220 nm Si for the upper taper in the final design maintains compatibility with standard CMOS foundry process flows, meaning less complex fabrication and an overall cost reduction. Finally, by using SiN, a material with a lower intrinsic refractive index, we establish a lower index contrast between waveguide and substrate on the lower chip which expands the mode size outside the waveguide (termed the evanescent tail). Through this modal expansion, we can improve the 1 dB lateral and vertical alignment tolerances compared to the coupling between two 220 nm silicon double tapers. Modal expansion is also the reason for the double linear taper design for both the upper and lower tapers - a long region with narrow lateral dimensions will result in a aerially larger evanescent field and thus allow for improved alignment tolerances over a single, linear taper. The goal of these couplers will be to increase lateral alignment tolerance beyond the capability of the self-aligning C4 or C2 bumps so that passive assembly for their lateral alignment can be accomplished. Vertically speaking, the goal will be to increase vertical alignment tolerance beyond the variation in bump height uniformity across a wafer or die since the height of the C2 bump after reflow will determine the vertical offset of the Si double taper and the SiN double taper (in other words, the gap between the upper and lower chip). Widening alignment tolerances to enable passive assembly of optical and electrical components directly equates to faster assembly speeds and lower costs. The remainder of this study will discuss how the simulations were setup and will show the final results demonstrating high coupling efficiency with widened alignment tolerances.

\section{Materials and Methods}\label{sec:sim_setup}
\subsubsection{Material Platform Constraints}
\par As part of the simulation setup, it is important to list the parameters, the material constraints, and the design variables to be calculated by the simulations. For our material constraints, we assume the upper waveguides are fabricated on standard 220 nm silicon on insulator (SOI) with a 2 $\mu$m bottom oxide (BOX) layer. This is consistent with the most popular material platform choice for PICs in high volume manufacturing (HVM). This immediately places an additional constraint on the upper film final output waveguide width (W$_{\text{Si}}$) because the widest single mode waveguide possible for 220 nm thick Si with $n$ = 3.45 is 440 nm. To simplify our initial simulations, we assume the BOX layer is thick enough such that none of the mode leaks into the upper chip's bulk silicon. In other words, for simulation purposes the upper substrate is effectively entirely silica with an index of 1.45. This assumption is reasonable for high to mid range index contrast waveguides such as those made from Si or SiN.

\par Next, we assume the lower waveguide layer is standard 220 nm SiN with refractive index of 2.0. The substrate below the SiN waveguiding layer is constrained to SiO$_2$ ($n$ = 1.45). This is because our broader goal is to develop vertical optical interconnects for use on glass optical interposer technology which has advantages in terms of thermal, mechanical, and electrical performance and makes alignment easier due to transparency \cite{2022brusberg}. It is assumed then that the input light to the waveguide comes from an edge coupled array of standard SMF-28 fibers operating at $\lambda = 1.55$ $\mu$m which are far away. In an effort to limit the longitudinal footprint of the coupler, we self-constrain our adiabatic taper to be no longer than 500 $\mu$m in length. Additionally, it is common in microelectronic packaging to use heat treated or UV curable epoxies as underfill materials in a flip chip bonded system. These underfill materials are typically formed from low-expansion fillers, such as fused silica, and a liquid prepolymer, such as a thermosetting resin for adhesion that can be cured to a solid composite and fill the gap between the chips via capillary action. The underfills serve the purpose of providing mechanical stability and improved reliability to the standard C4 or C2 interconnects through thermal mismatch reduction and hermetic sealing \cite{2016lau}. In addition to thermal expansion coefficient matching, these underfills can also provide refractive index matching by have tunable refractive indices from 1.45-1.6 \cite{2013epotek, 2015AMS}. We assume in this study we are using an underfill such as EPO-TEK$^{\text{\textregistered}}$ 305 which has a refractive index of 1.45 at 1550 nm, instead of an air gap with refractive index of 1. This constraint will later be relaxed and the effect of a small variation of the underfill refractive index on coupling efficiency will be demonstrated (between 1.4 and 1.6). Finally, the input waveguide width (W$_{\text{SiN}}$) will be set to 1 $\mu$m as this provides an input mode with a high confinement factor to the coupler.

\begin{table}[ht]
\centering
\begin{tabular*}{\textwidth}{c|c|c}
\hline
& \multicolumn{2}{c}{\textbf{Platform Constraints}}\\
\hline
\textbf{Parameters} & \textbf{Upper taper} & \textbf{Lower taper}\\
\hline
Material & Si & SiN \\
\hline
Refractive index & 3.45 & 2.0 \\ 
\hline
Thickness (t) &  \multicolumn{2}{c}{220 nm}\\
\hline
Waveguide width (W$_{\text{Si}}$ and W$_{\text{SiN}}$) & 440 nm & 1 $\mu$m\\
\hline
Tip width (W$_{\text{tip, Si}}$ and W$_{\text{tip, SiN}}$ ) & \multicolumn{2}{c}{$>$ 100 nm (193 nm immersion lithography resolution \cite{2019fahrenkopf})}\\
\hline
Non-adiabatic taper length ($\ell$) & \multicolumn{2}{c}{10 $\mu$m}\\
\hline
Taper slope profile & \multicolumn{2}{c}{all tapers are linear}\\
\hline
Adiabatic taper length (L) & \multicolumn{2}{c}{$<$ 500 $\mu$m}\\
\hline
Substrate material & 2 $\mu$m BOX atop bulk Si & SiO$_2$\\
\hline
Inter-chip gap material & \multicolumn{2}{c}{UV curable or heat treatable underfill epoxy}\\
\hline
Underfill refractive index & \multicolumn{2}{c}{1.45}\\
\hline
& \multicolumn{2}{c}{\textbf{Design Variables}}\\
\hline
\textbf{Parameters} & \textbf{Lower Limit} & \textbf{Upper Limit}\\
\hline
Si Intermediate width (W$_{\text{Si, i}}$) & 200 nm & 250 nm\\
\hline
SiN Intermediate width (W$_{\text{SiN, i}}$) & 600 nm & 700 nm\\
\hline
Si tip width (W$_{\text{tip, Si}}$) & 70 nm & 150 nm\\
\hline
SiN tip width (W$_{\text{tip, SiN}}$) & 100 nm & 250 nm\\
\hline
\end{tabular*}
\caption{Summary of parameters, material constraints, and design variables. Note that the ranges of 200-250 nm and 600-700 nm for intermediate width study were selected following preliminary 3D eigenmode expansion simulations which originally varied from 170-440 nm and 500-1000 nm for Si and SiN, respectively. Also, a point of minor importance is that the Si tip width was varied below 100 nm (down to 70 nm) in order to help determine fabrication error tolerance.} 
\label{tab:parameters}
\end{table}

\par Furthermore, in order to maintain simplicity of fabrication, we constrain all taper slope profiles to be linear instead of nonlinear profiles such as a quadratic, cubic, quartic, or exponential slope. As was shown in \cite{2017itoh}, a non-linear slope actually shrinks the necessary taper length necessary to achieve a given coupling efficiency; however, non-linear profiles require lithographic resolution which may be higher than standard 193 nm DUV or immersion techniques can provide. The double taper design attempts to mimic the non-linear profiles, but maintain linear segments which relax such lithographic requirements. 
 
As a result, the only parameters left to vary are the upper and lower taper tip widths, upper and lower non-adiabatic taper lengths, and the intermediate waveguide widths. However, because the non-adiabatic lengths convert the mode size from one size to another within the the same SiN or Si thin film with no changes in thickness or background index, these widths can be set to 10 $\mu$m as in \cite{2016itoh} as they have little influence on the overall performance of the out of plane coupler. For the tip widths of the upper and lower tapers, which are anticipated to be the smallest feature size of the design, we assume a lower limit of 100 nm for both the Si and SiN tapers. This allows our design to remain above the resolution capabilities of industry standard 248 nm DUV, 193 nm DUV, and 193 nm immersion lithography techniques for Si-PICs utilizing thin/thick SOI and bulk Si according to \cite{2020IPSRSiP, 2019fahrenkopf}. The material constraints and remaining design variables are summarized in Table \ref{tab:parameters} .

\subsubsection{3D FDTD Simulation Setup}
\par The performance metrics of interest in this study are insertion loss (in units of dB), 1 dB lateral alignment tolerance (in units of $\mu$m), 1 dB vertical alignment tolerance (in units of $\mu$m), 1 dB twist and tilt alignment tolerance (in units of degrees, with twist indicating $+\phi$ rotation and tilt indicating $+\theta$ rotation as defined in Figure \ref{fig:crosstapers}), and 1 dB bandwidth (in units of nm). In order to calculate these outputs we will primarily use Lumerical's 3D Finite Difference Time Domain (FDTD) tool. The use of a 3D FDTD solver to simulate performance is important because it is able to take into account loss due to scattering out of plane, which is likely to occur at the tips of the tapers. 

\par For the 3D FDTD simulation setup, meshes which were one-fourth the smallest feature size were used over the SiN and Si regions (this means they varied from 15-40 nm depending on the value of W$_{\text{tip, Si}}$ and W$_{\text{tip, SiN}}$ to optimize simulation time). The background mesh for the solver was set to Lumerical's preset auto-non uniform mesh with a mesh accuracy of 2 (this equates to approximately a 50 nm background mesh for a wavelength of 1.55 $\mu$m). In terms of boundary conditions, perfectly matched layer (PML) boundary conditions, which assume absorption of incident electric field, were used to account for loss due to out of plane scattering of light. A monochromatic TE mode source with a wavelength of 1.55 $\mu$m was used and transmission was measured through an output monitor 2 $\mu$m by 2 $\mu$m in size located 12 $\mu$m past the end of the final, non-adiabatic taper in the Si. For the 1 dB bandwidth calculations, a broadband light source was used with 300 nm bandwidth extending from 1.3 $\mu$m to 1.6$\mu$m, and the system was simulated with a TE mode and a TM mode separately to determine polarization dependent losses. 
 
\par Vertical offset sweeps were performed where the inter-chip gap was varied from 0 to 3 $\mu$m for intermediate waveguide widths between 200-250 nm for Si and 600-700 nm for SiN to determine the effect of intermediate width on vertical alignment tolerance. For the vertical offset sweep, a 0 $\mu$m gap indicates the Si and SiN tapers are directly on top of one another and a 3 $\mu$m gap indicates the distance from the top of the SiN taper to the lowest point of the Si taper. We assume, since the intensity of optical modes in waveguides follows a Gaussian distribution, that the trends which prevail for vertical offset sweeps will prevail for lateral offset sweeps (and thus lateral alignment tolerance). In other words, only vertical offset sweeps were performed for each intermediate width value, the trends were extracted, and then lateral offset simulations were completed later with the optimized widths to determine the final lateral alignment tolerance (and save time in simulating the structure).
The effect of symmetric and asymmetric tip widths was also investigated by first varying the tip width of both the Si and SiN identically, and then varying only the Si while the SiN tip width remained constant at 100 nm and vice versa. To properly assess the effect on maximum coupling efficiency and alignment tolerance, vertical offset sweeps were performed identically to the intermediate width case.
\par Next, with the intermediate widths and tip widths finalized, the translational and angular alignment tolerances were characterized. This was done by performing vertical offset sweeps from 0 to 3 $\mu$m as discussed, a lateral offset sweep where $z$ and $x$ offset were set to zero and $y$ offset was varied from 0 to 3 $\mu$m, and a longitudinal offset sweep where $y$ and $z$ offsets were zero and $x$ offset was varied from 0 to 150 $\mu$m. Likewise, the angular alignment tolerances were simulated by instituting twists ranging from 0 to 3 degrees and tilts ranging from 0 to 0.5 degrees of the Si double taper with respect to the SiN double taper (which remained at a 0 degree twist and tilt rotation at all times). 
\par Finally, variation in the refractive index of the underfill region was conducted by changing the underfill index from 1.4 to 1.6 (maintaining the 50 nm background mesh) and simulating the coupling efficiency. 
Now, the results of these simulations will be presented in the following section. 

\begin{figure}[htbp]
\centering
\subfigure[\label{subfig:si_width_vary_3dfdtd}]{\includegraphics[scale=0.215]{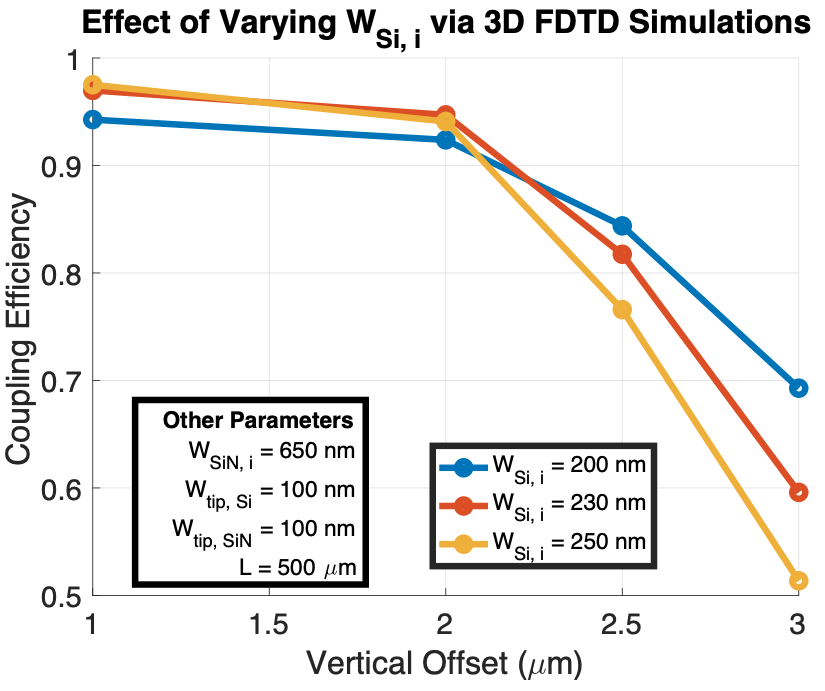}}
\subfigure[\label{subfig:sin_width_vary_3dfdtd}]{\includegraphics[scale=0.175]{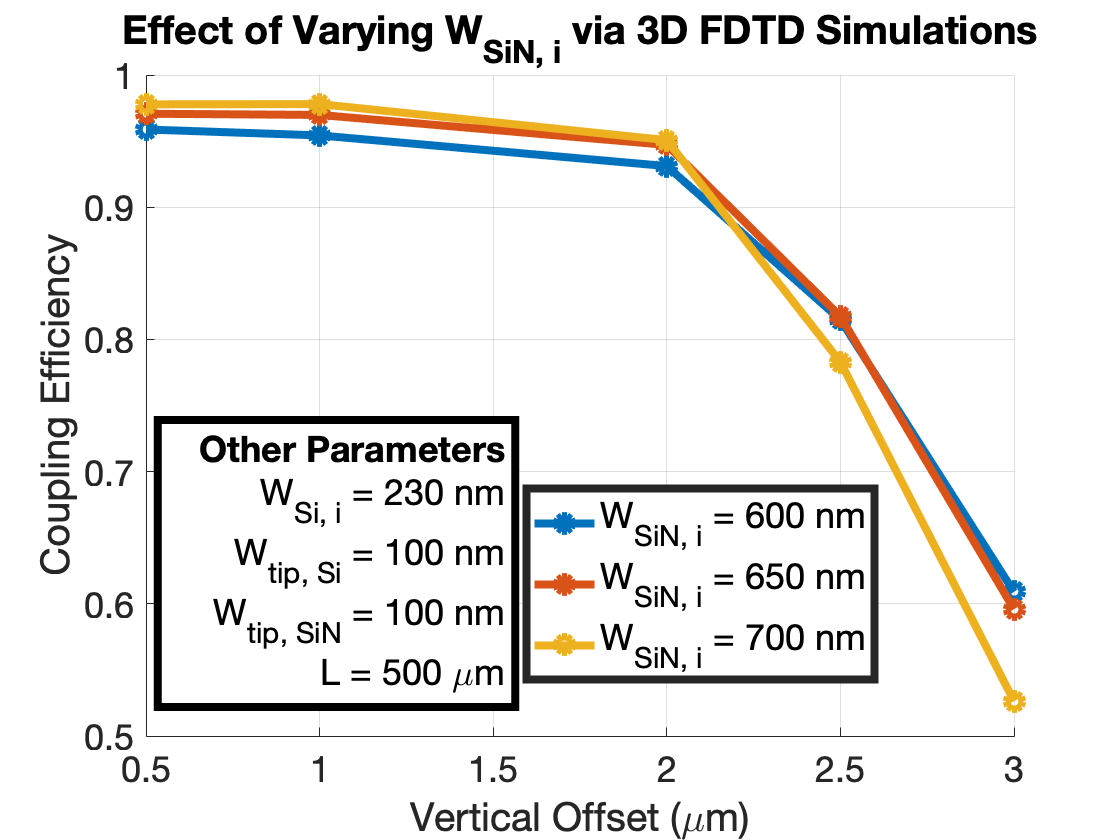}}
\subfigure[\label{subfig:si_width_side_profile_200}]{\includegraphics[scale=0.38]{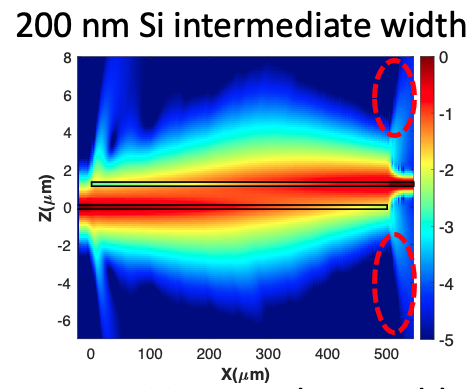}}
\subfigure[\label{subfig:si_width_side_profile_250}]{\includegraphics[scale=0.4]{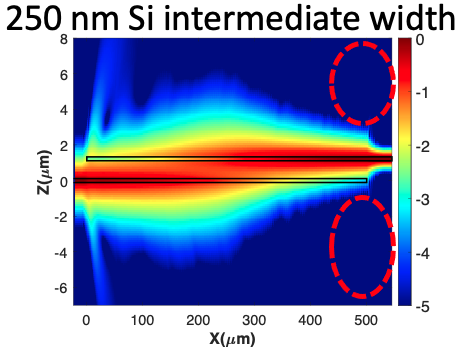}}
\caption{The effect of a small variation in Si and SiN intermediate width on coupling efficiency is shown following 3D FDTD simulations. In \subref{subfig:si_width_vary_3dfdtd} and \subref{subfig:sin_width_vary_3dfdtd}, we see that decreasing the Si or SiN intermediate width decreases maximum coupling efficiency while improving alignment tolerance (and vice versa). We also see that the structure is relatively robust to small variations of $\pm$ 25-30 nm around 230 nm and $\pm$ 50 nm around 650 nm which could arise due to process variation. In \subref{subfig:si_width_side_profile_200}, we see a cross sectional plot of the optical mode intensity on a logarithmic scale. The red dotted circles are to draw the readers attention. Notice that for the case of the 200 nm Si intermediate width, there is increased out of plane scattering at the point when the SiN and Si tapers stop overlapping. This scattering is shown by the lighter shade of blue in the red dotted circles. Moreover, this out of plane scattering is significantly reduced in the case of the 250 nm wide Si, shown by the dark shade of blue in the red dotted circles of \subref{subfig:si_width_side_profile_250} (because the dark blue region represents -50 dB of loss and the lighter blue represents approximately -30 dB of loss, this difference in color is important).}
\label{fig:si_width_vary_3dfdtd}
\end{figure}

\section{Results}
\par Using the simulation setup described in Section \ref{sec:sim_setup}, the results showing the effect of intermediate width variation are presented in Figure \ref{fig:si_width_vary_3dfdtd} for both Si and SiN. From the data in \ref{subfig:si_width_vary_3dfdtd}, we find that an Si intermediate width of 230 nm provides maximum coupling efficiency above 95\% with vertical alignment tolerance beyond 2.5 $\mu$m.  In Figure \ref{subfig:sin_width_vary_3dfdtd}, we find a similar result for an SiN intermediate width of 650 nm. The plots together show an important trend - a decrease in intermediate width results in widened alignment tolerances at the expense of maximum coupling efficiency. Due to this finding, we set the Si and SiN intermediate widths to 650 nm and 230 nm for the remainder of the study.
\par
\begin{figure}[htbp]
\centering
\subfigure[\label{subfig:symmetric_tip_width_vary}]{\includegraphics[scale=0.21]{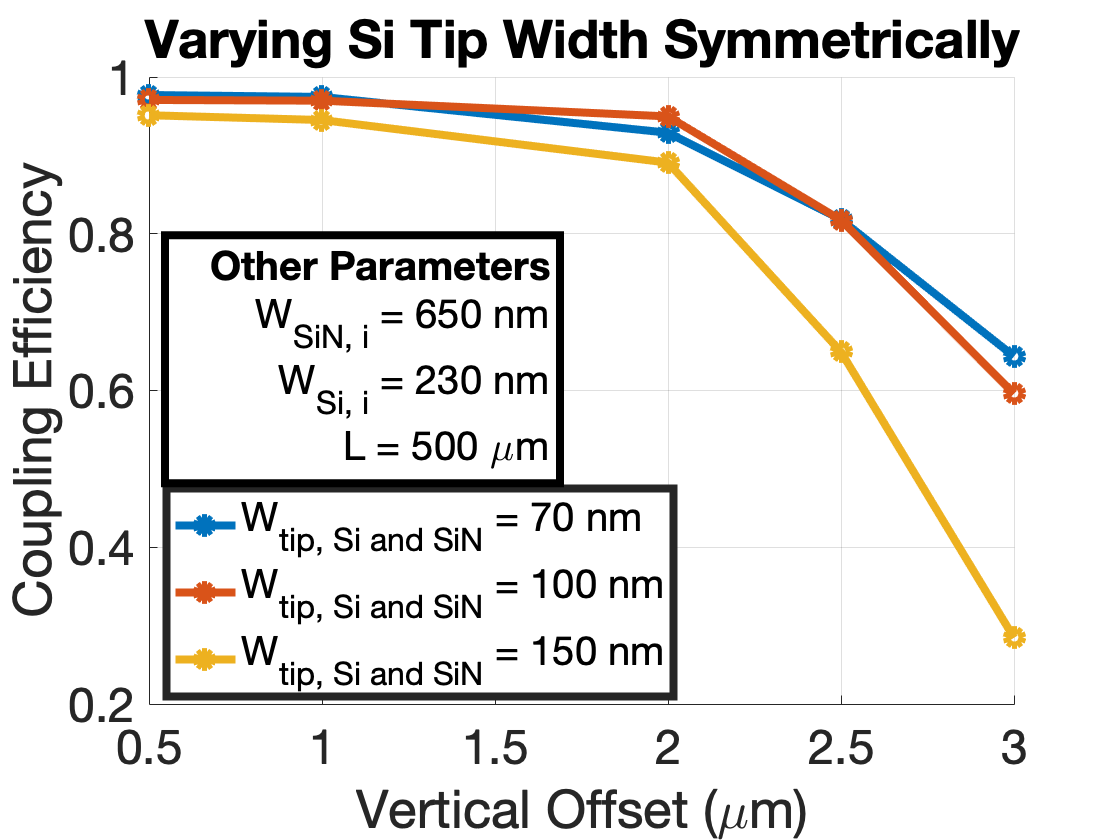}}
\subfigure[\label{subfig:tip_width_mode}]{\includegraphics[scale=0.4]{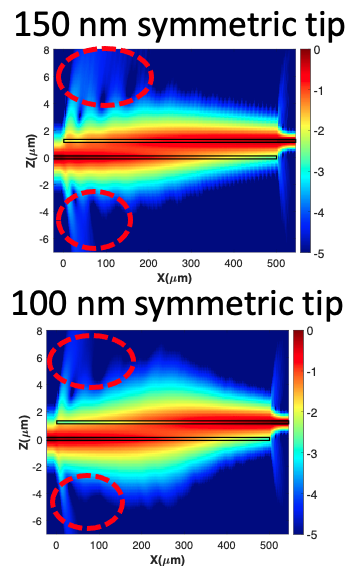}}
\subfigure[\label{subfig:sin_tip_width_vary}]{\includegraphics[scale=0.165]{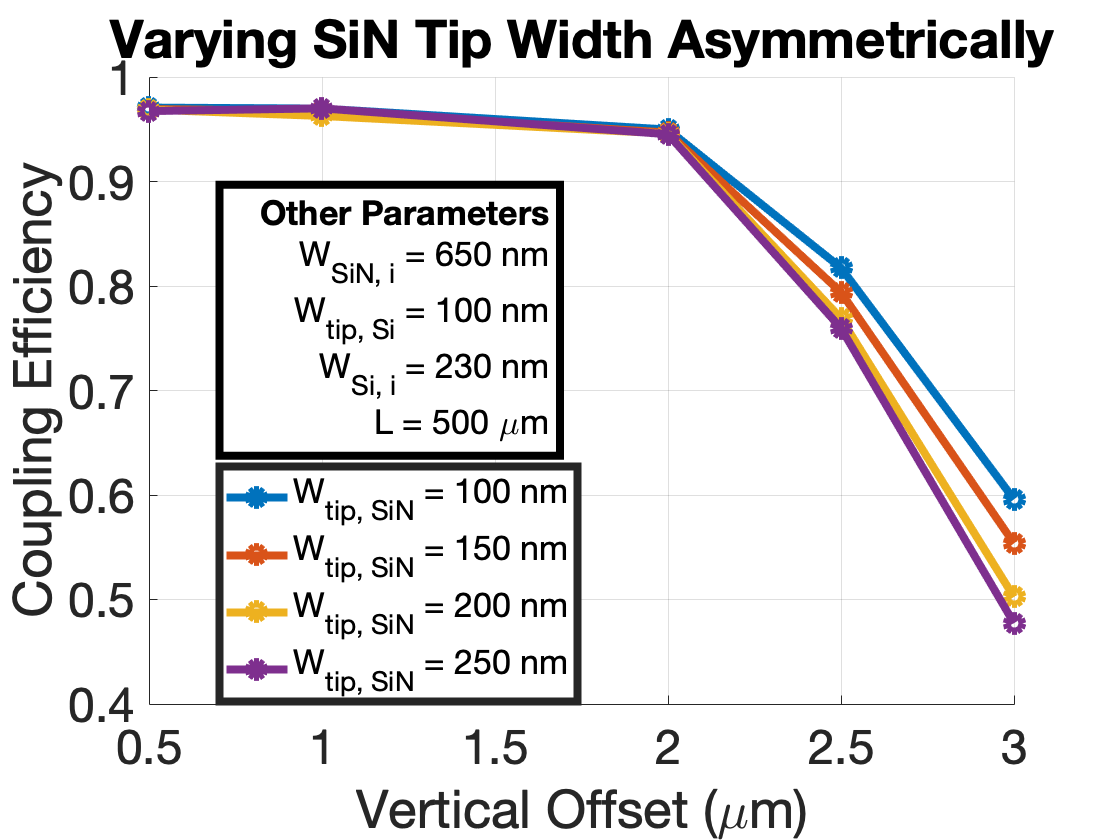}}
\subfigure[\label{subfig:si_tip_width_vary}]{\includegraphics[scale=0.165]{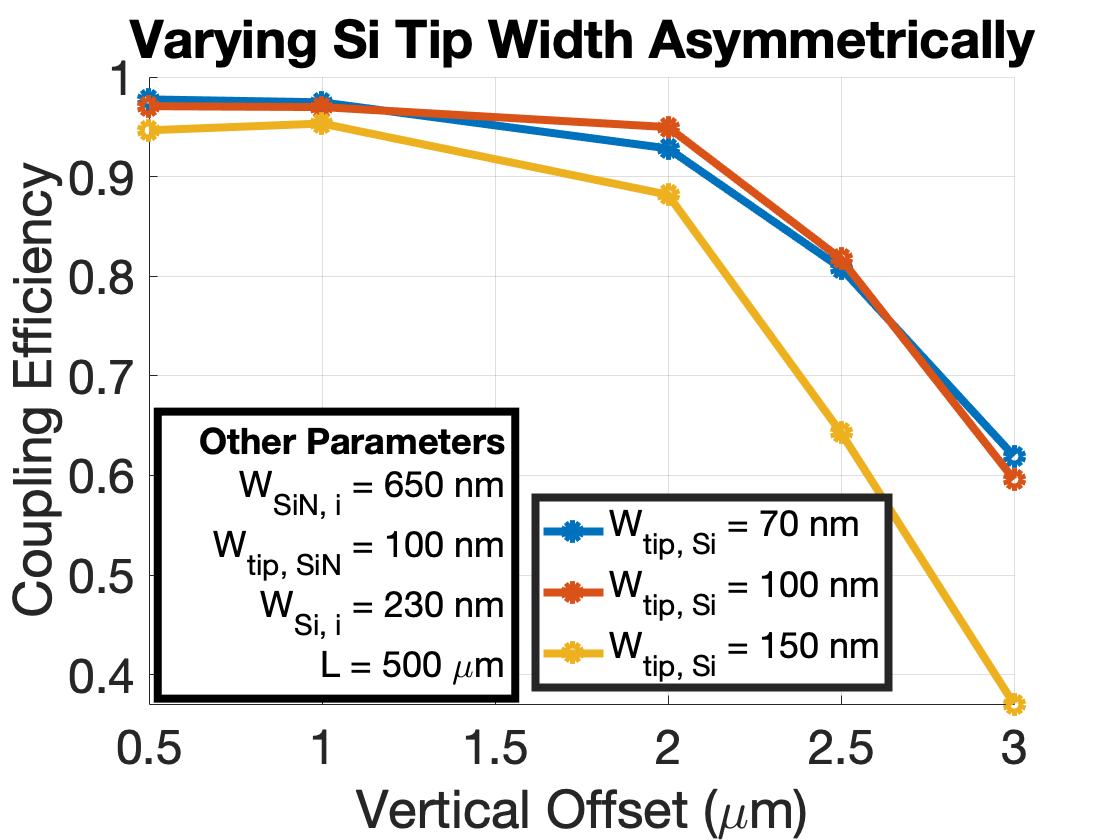}}
\caption{The effect of tip width on coupling efficiency is depicted. First, in \subref{subfig:symmetric_tip_width_vary}, tip width is symmetric (the same for Si and SiN double tapers) and variation shows that increasing tip width decreases coupling efficiency and alignment tolerance. In \subref{subfig:tip_width_mode}, a cross section of the optical mode intensity on a logarithmic scale is shown for symmetric tip widths of 100 nm and 150 nm, with red dotted circles to draw the reader's attention. The important takeaway is that when going from the 100 nm plot to the to 150 nm plot, the red dotted circle contains a lighter shade of blue, indicating out of plane scattering losses as a result of a wider tip width. In \subref{subfig:sin_tip_width_vary}, tip width is asymmetric (different for Si and SiN double tapers) and variation of the SiN tip width while Si is held at 100 nm shows that increasing tip width decreases alignment tolerance with a negligible impact on maximum coupling efficiency. In \subref{subfig:si_tip_width_vary}, again tip width is asymmetric and variation of Si tip width shows that increasing tip width beyond approximately 100 nm decreases coupling efficiency and alignment tolerance. All three data figures demonstrate the coupler is relatively robust to small variation in tip width (the smallest feature size of the design due to process variation). }
\label{fig:tip_width_vary}
\end{figure}

 Next, we present the data showing the effect of asymmetric tip widths and a small variation in symmetric tip width, as shown in Figure \ref{fig:tip_width_vary}. From Figure \ref{subfig:symmetric_tip_width_vary}, it is clear that increasing the tip widths symmetrically decreases both maximum coupling efficiency and alignment tolerance. From Figure \ref{subfig:sin_tip_width_vary}, we conclude that increasing SiN tip width asymmetrically (while Si tip width is constant at 100 nm) decreases the alignment tolerance while having a negligible effect on maximum coupling efficiency. From Figure \ref{subfig:si_tip_width_vary}, we conclude that increasing Si tip width asymmetrically (while SiN tip width is constant at 100 nm) decreases both maximum coupling efficiency and alignment tolerance, while decreasing the Si tip width below 100 nm provides marginal improvement. As a result of these findings, we conclude a symmetric tip width of 100 nm provides for widened alignment tolerances while maintaining high maximum coupling efficiency.

The final values for the design variables are summarized in Table \ref{tab:finalparameters}.
\par
\begin{figure}[htbp]
\centering
\subfigure[\label{subfig:xyz_vary}]{\includegraphics[scale=0.4]{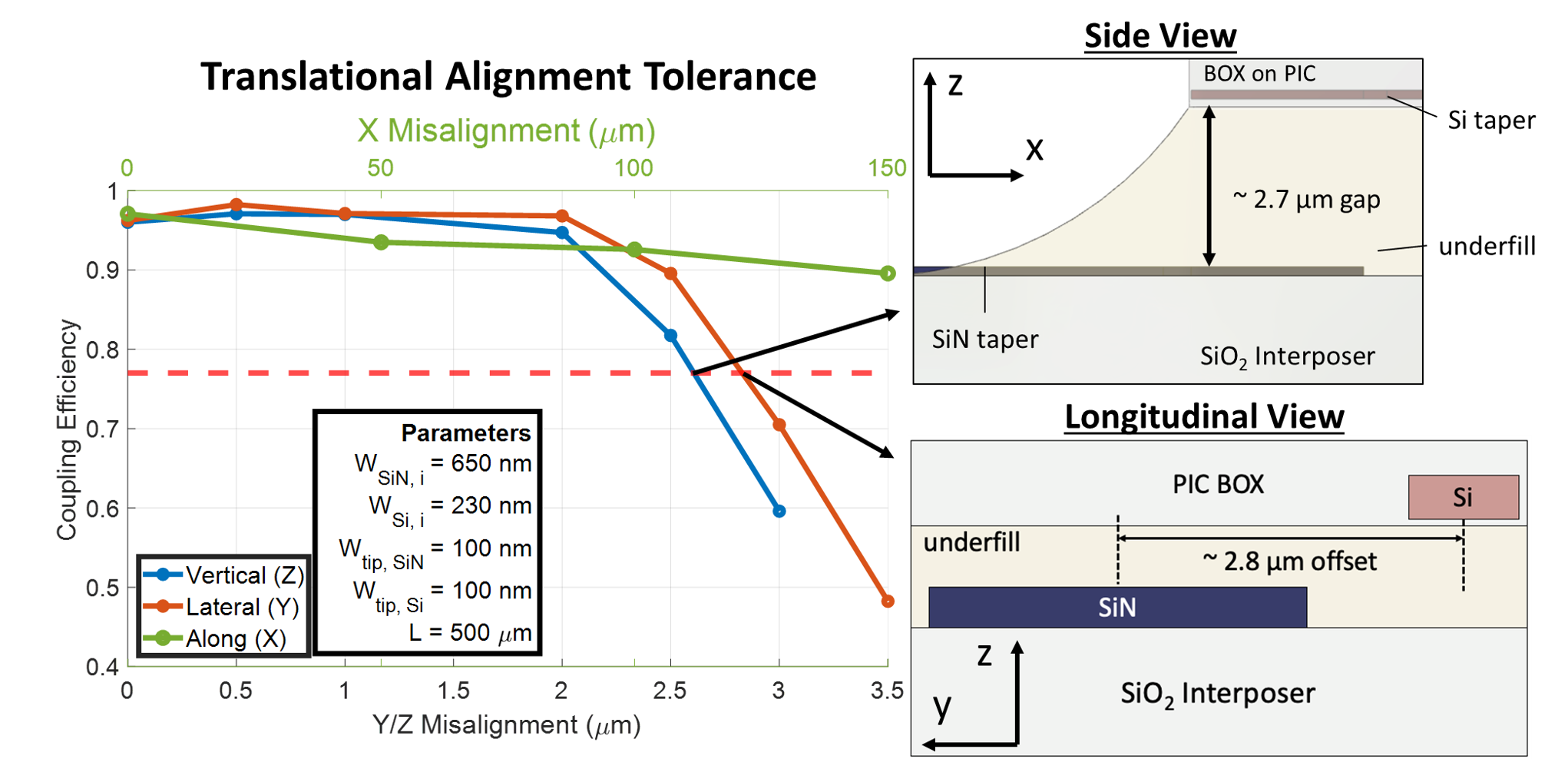}}
\subfigure[\label{subfig:rotation_vary}]{\includegraphics[scale=0.4]{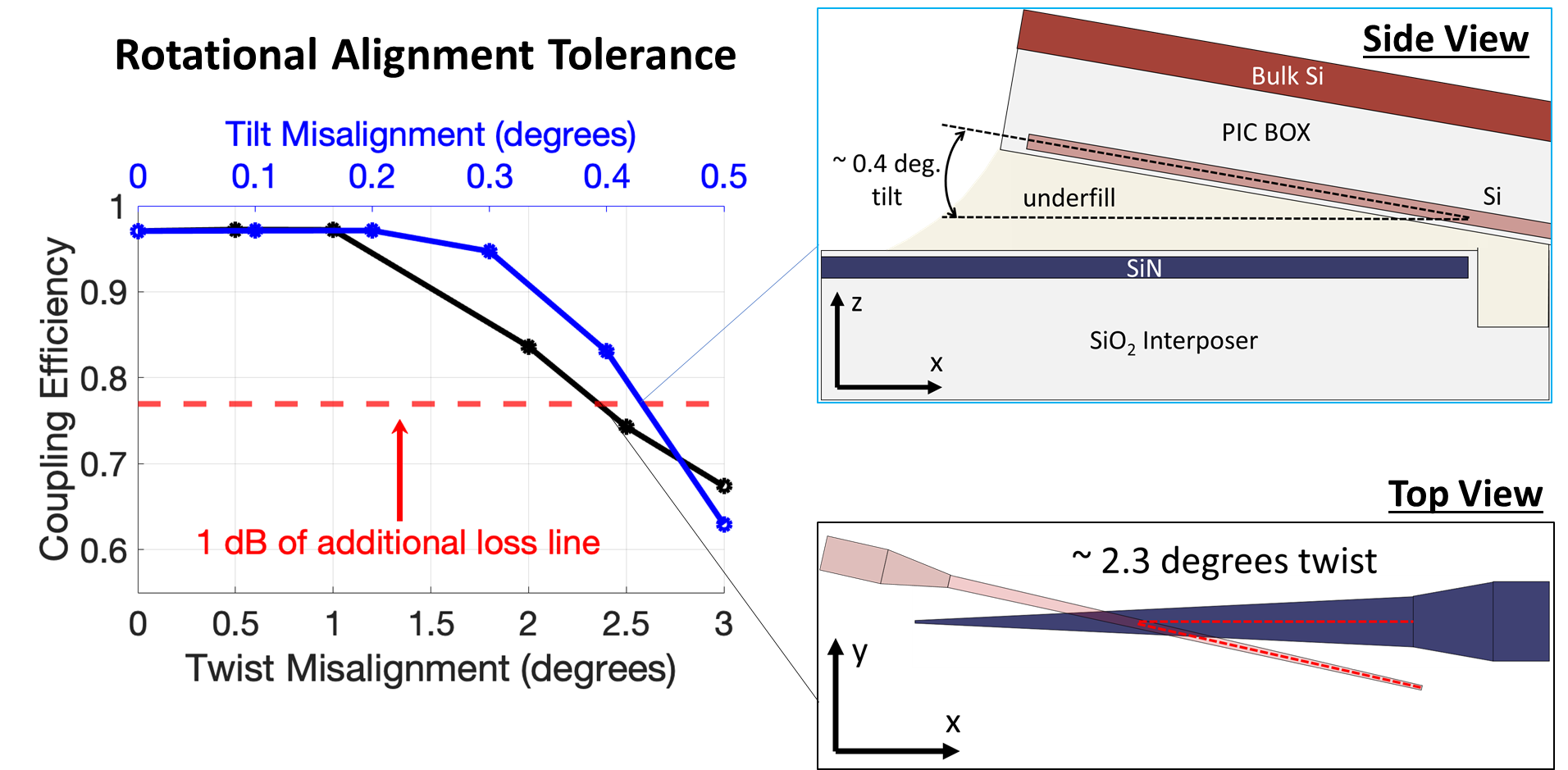}}
\caption{The final translational and rotational alignment tolerances using 3D FDTD simulations are shown. In \subref{subfig:xyz_vary}, a plot of coupling efficiency versus misalignment in $x$, $y$, and $z$. Note that while simulating the $z$ alignment tolerance, the $y$ and $x$ offsets were set to zero and while simulating the $y$ and $x$ alignment tolerance, the $z$ offset was set to zero (so there was no gap between the tapers). The red dotted line on the figure represents 1 dB of additional loss when beginning from 97\% (or 0.13 dB) of insertion loss (meaning the red dotted line is at 1.13 dB of loss or 77 \% coupling efficiency). From the plot, we conclude that the final 1 dB vertical alignment tolerance is approximately 2.7 $\mu$m, the 1 dB lateral ($y$) alignment tolerance is approximately $\pm$ 2.8 $\mu$m, and the 1 db longitudinal (or ``along") alignment tolerance is greater than 150 $\mu$m. In \subref{subfig:rotation_vary}, a similar plot is shown for twist misalignment (rotation about the $z$ axis) and tilt misalignment (rotation about the $y$ axis). We conclude that the final 1 dB twist alignment tolerance is approximately 2.3 degrees and the 1 dB tilt alignment tolerance is approximately 0.4 degrees. The arrows in \subref{subfig:xyz_vary} and \subref{subfig:rotation_vary} leading from the data to the side view, top view, and longitudinal view show exactly where the tapers are relative to one another when the alignment tolerance limits are reached. In the diagrams on the right side, coordinate systems are presented to orient the reader. Final design parameters are also shown in the plot for completeness.}
\label{fig:alignment_tolerance}
\end{figure}

\begin{figure}[htbp]
\centering
\subfigure[\label{subfig:underfill_vary}]{\includegraphics[scale=0.17]{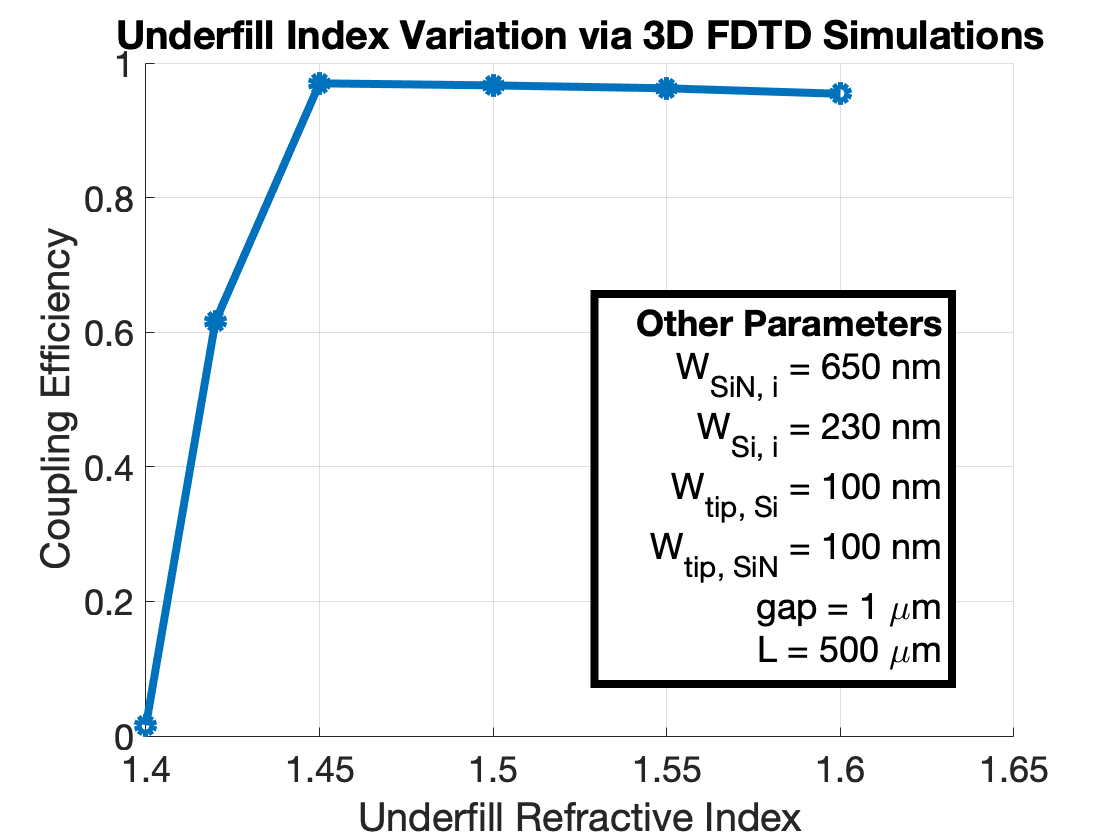}}
\subfigure[\label{subfig:bandwidth}]{\includegraphics[scale=0.165]{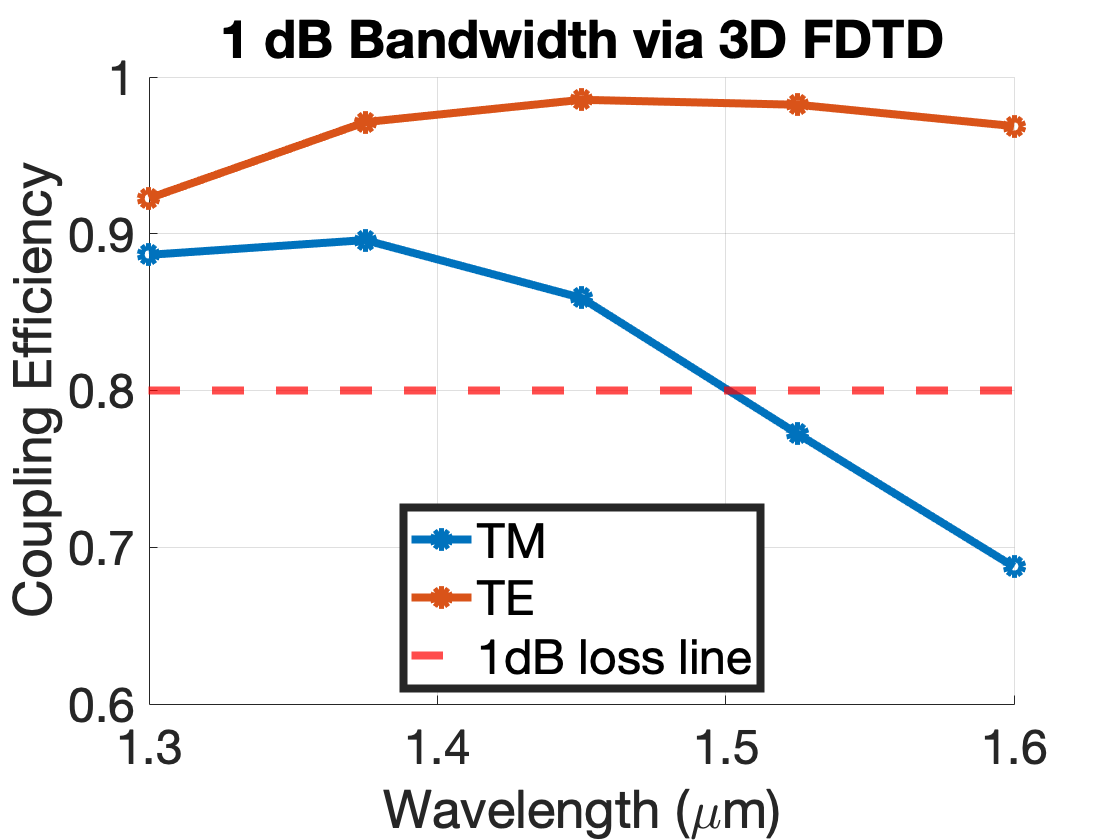}}
\caption{The effect of the underfill refractive index, wavelength, and polarization are shown. In \subref{subfig:underfill_vary}, the underfill refractive index is varied for a 1 $\mu$m gap is shown. Notice that between 1.45 and 1.6, the typical range of refractive indices for UV or heat treatable optical underfill epoxies, the coupling efficiency is relatively unaffected. On the other hand, going below 1.45 for the underfill refractive index causes the optical mode to become more confined in the higher index waveguide-substrate system and significantly affects coupling. In other words, this indicates that having an air gap between the two chips with refractive index $n$ = 1 would eliminate coupling entirely. In \subref{subfig:bandwidth}, the wavelength is varied between 1300 nm and 1600 nm for the transverse electric (TE) and transverse magnetic (TM) modes. From this plot, we conclude the 1 dB bandwidth for TE mode propagation is greater than 300 nm, the polarization dependent losses at 1550 nm are approximately 1 dB and are approximately 0.2 dB at 1310 nm.}
\label{fig:underfill_vary}
\end{figure}

 Moreover, the 1 dB translational and rotational alignment tolerance data can be found in Figure \ref{fig:alignment_tolerance} and the 1 dB bandwidth data can be found in Figure \ref{subfig:bandwidth}. From the data in Figure \ref{subfig:xyz_vary}, the 1 dB $x$, $y$, and $z$ alignment tolerances were determined to be $>$ 150 $\mu$m, $\pm$ 2.8 $\mu$m, and 2.7 $\mu$m, respectively. From the data in Figure \ref{subfig:rotation_vary}, the 1 dB twist and tilt alignment tolerances were determined to be approximately 2.3 degrees and 0.4 degrees, respectively. Finally, from the data in Figure \ref{subfig:bandwidth} we obtain two pieces of information: the 1 dB bandwidth for TE mode coupling is greater than 300 nm and the polarization dependent losses for 1550 nm injection are greater than 1 dB of loss (20-25\%). 
\par The final set of data shown in Figure \ref{subfig:underfill_vary} shows the effect of the underfill refractive index. For an underfill with refractive index between 1.45 and 1.6, the coupling efficiency is relatively unchanged (less than 5\% shift). On the other hand, going below 1.45 results in significant losses greater than 1 dB. This result indicates an air gap would be insufficient for TE mode coupling at a 1 $\mu$m inter-chip spacing.

\begin{table}
\centering
\begin{tabular}{c|c}
\hline
\textbf{Parameters} & \textbf{Final Values for Design Variables}\\
\hline
Si Intermediate width (W$_{\text{Si, i}}$) & 230 nm\\
\hline
SiN Intermediate width (W$_{\text{SiN, i}}$) & 650 nm\\
\hline
Si tip width (W$_{\text{tip, Si}}$) & 100 nm\\
\hline
SiN tip width (W$_{\text{tip, SiN}}$) & 100 nm\\
\hline
\end{tabular}
\caption{Summary of the final values for the design variables of this study following 3D FDTD simulation optimization.} 
\label{tab:finalparameters}
\end{table}

\section{Discussion}
\par From the results in the prior section, we propose a few underlying mechanisms which can used for future design of vertical evanescent couplers. First, we propose that the decrease in maximum coupling efficiency as W$_{\text{Si, i}}$ decreases below 230 nm is due to repeated polarization conversion.
 
In general, below a 220 nm Si waveguide width for 220 nm thick Si, the mode with the highest effective refractive index is the TM mode. Therefore, in coupling to a waveguide with W$_{\text{Si, i}}$ = 220 nm, polarization conversion must occur twice leading to increased losses (once going from TE to TM during coupling to the Si and once going from TM to TE in the final, short Si taper). Going forward, we will limit the Si intermediate width to 230 nm to avoid these losses.
\par The second important mechanism to elaborate on is the cause of the loss seen in Figure \ref{subfig:tip_width_mode} when going from the 100 nm tip width to the 150 nm tip width. This loss is the result of out of plane scattering due the mode interacting with a high index Si region during propagation - a narrower tip width results in decreased out of plane scattering and a higher maximum coupling efficiency. In terms of alignment tolerance, a narrower tip width increases the slope of the adiabatic taper and makes the entire taper thinner, resulting in modal expansion and thus improved coupling for larger $z$ or $y$ offsets. 
\par One point to highlight about the results in the prior section, is that the plots also demonstrate the robustness of the double taper structure to variation in tip width, intermediate width, and underfill refractive index due to natural process variation during fabrication. For example, even with an error of 50 nm in tip width for Si, or 150 nm in the tip width of SiN, the effect on maximum coupling efficiency is less than 5 \% and on 1 dB vertical alignment tolerance is less than 500 nm. This assumes that the variation occurs between 1.45 and 1.6 for the underfill refractive index - the result of Figure \ref{subfig:underfill_vary} indicates an air gap would be insufficient for TE mode coupling at a 1 $\mu$m inter-chip spacing, an important caveat.
\subsection{Comparison}
\par A brief review of the performance metrics for alternate vertical couplers are presented in Table \ref{tab:review}. In comparing the values for insertion loss, alignment tolerance, and bandwidth associated with the double taper structure from this study, our coupler stacks up well to peer innovations. First, in terms of insertion loss, our vertical cross tapers display a maximum coupling efficiency of slightly greater than 97\%, or 0.13 dB of loss. This is competitive compared to the other high performance couplers which have insertion losses of 0.45 dB or less (90\% coupling efficiency or more). Second, our widened translational alignment tolerances above 2.7 $\mu$m compare favorably to other evanescent, edge, or grating coupler designs which are typically less than 2 $\mu$m. One important point to keep in mind when comparing our alignment tolerances to a free form coupler, which can achieve exceptionally large vertical alignment tolerances, is the monolithic, planar nature of our design which relies on standard, controlled foundry processes and CMOS compatible materials. This is contrasted by couplers which are fabricated using two-photon polymerization techniques, a process with significantly lower throughput due to the sequential patterning of optical components with repeated exposures instead of processing devices in parallel using batch patterning processes. In terms of material properties, it also means our coupler is reflow temperature compatible and not subject to the same mechanical, thermal, or humidity based reliability problems which may arise for polymer based structures. Additionally, because the evanescent coupler in this study maintains mode diameters less than 10 $\mu$m, even during the modal expansion occurring within the tapers, compared to expanding the mode to hundreds of microns in diameter as is routine in free form couplers, our lateral pitch can be made significantly finer. A finer I/O pitch directly equates to a higher I/O density along the shoreline of the PIC or TxRx, and thus a higher possible bandwidth. Thus, our coupler may provide more rapid, less costly scaling to HVM applications requiring high density integration which is especially crucial in the context of scaling to greater than 1 Pbps datacom or telecom ToR switching packages. 

\definecolor{LightRed}{rgb}{0.95 0.85 0.85}
\begin{table}[ht]
\footnotesize
\centering
\renewcommand{\arraystretch}{1.2}
\newcolumntype{C}[1]{>{\centering\arraybackslash}m{#1}}
\footnotesize
\begin{tabular*}{\textwidth}{C{2.1cm}|C{1.5cm}|C{1.9cm}|C{2cm}|C{1.5cm}|C{1.7cm}}
\hline
\textbf{Package} & \textbf{Coupler} & \textbf{IL$^1$ (dB)} & \textbf{1dB TOL$^2$} \textbf{($\mu$m)} & \textbf{1dB BW$^3$} \textbf{(nm)} & \textbf{Year and Ref.}\\ 
\hline
VCSEL to Si-PIC &  Grating & -11.8 & $\pm$ 1.6 & $<$ 50 & 2016 \cite{2016obrien} \\
\hline
InP laser to Si-PIC& Edge & -1.3 (TE) \newline -1.5 (TM) & $< \pm$ 2 (lateral) & $>$ 100 & 2016 \cite{2016barwicz, 2015nah, 2016martin} \\ 
\hline
Si-PIC to Si-PIC& Grating & -4.0 & $\pm$ 1.25 (lateral) & $<$ 10 &2017 \cite{2017zonou} \\
\hline
InP, SMF, Si-PIC, to Si-PIC & Photonic Wire Bond & -0.4 to -1.3 (TE)  & \textbackslash & $\geq$ 50 & 2018 \cite{2018billah}\\
\hline
Si-PIC to Polymer-interposer& Evanescent & -2 (TE)  & $\pm$ 2 (lateral) & $\geq$ 300 & 2018 \cite{2018dangel}\\
\hline
Si-PIC to Si-PIC& Free Form & -0.22 (TE)\newline -0.25(TM) & $>$ 35 (vertical)\newline $\pm$ 1.3 (lateral)& $>$ 300 &2020 \cite{2020shaoliang}\\
\hline
Polymer laminate to Si-PIC & Evanescent & -0.2 & > 5 (lateral)\newline 0.5 (vertical)& 180 &2021 \cite{2021englund}\\
\rowcolor{LightRed}
\hline
\textbf{SiO$_2$ interposer to Si-PIC} & \textbf{Evanescent} & \textbf{-0.13} (TE) & $\pm$\textbf{ 2.8 (lateral)}\newline \textbf{ 2.7 (vertical)} \newline \textbf{500 (along)} \newline \textbf{2.5 degree (twist)} \newline \textbf{0.4 degree (tilt)} & $>$ \textbf{300} & \textbf{2022 (this paper)}\\
\hline
\end{tabular*}
\vskip1em
\begin{flushleft}
$^1$IL = insertion loss (maximum coupling efficiency)\\
$^2$TOL = misalignment tolerance\\
$^3$BW = bandwidth
\end{flushleft}
\caption{Summary of a few relatively recent examples of chip-to-chip optical coupling schemes. Note that this list is not all encompassing and is meant to yield a general idea of competing designs and performance.} 
\label{tab:review}
\end{table}

\section{Conclusion}
\label{sec:conc}
In retrospect, this paper discussed the evanescent coupling of light from chip to chip using overlapping, inverse double tapers. Simulations using Lumerical's 3D FDTD solver demonstrated how 1 dB translational and rotational alignment tolerances for chip to chip coupling can be significantly expanded by using SiN, a low index contrast material platform, for the lower double taper. Moreover, because the lateral alignment tolerances of the coupler are approximately 2.8 $\mu$m, passive self-assembly can be accomplished through either reflow self-alignment using standard C4 solder bumps or Cu micro-pillar arrays thereby significantly reducing packaging costs. In addition, our coupler maintains a monolithic, planar design compatible with standard, well controlled CMOS foundry processes without the need to two photon polymerization, thereby reducing cost further and making the coupler an excellent candidate for use in high volume manufacturing applications. By achieving a fine interconnect pitch of < 10 $\mu$m relative to other vertical optical coupling techniques, this design increases the possible interconnect density along the shoreline of the PIC. This improvement in interconnect density directly correlates with improvements in bandwidth capacity, energy efficiency, and cost per interconnection. Given these facts, we conclude this vertical optical interconenct provides an enabling technology for future co-packaged designs to be used in applications such as hyperscale datacenters, self-driving automobiles, RF devices, and sensing for environmental and biological monitoring.

\bibliography{references}
\end{document}